\documentclass[reprint,aip,apl]{revtex4-1}

\usepackage{graphicx}
\usepackage{epstopdf}
\usepackage{dcolumn}
\usepackage{bm}

\begin{document}
\title{Single-charge detection by an atomic precision tunnel junction}

\begin{abstract}
We demonstrate sensitive detection of single charges using a planar tunnel junction 8.5\,nm wide and 17.2\,nm long defined by an atomically precise phosphorus doping profile in silicon. The conductance of the junction responds to a nearby gate potential and also to changes in the charge state of a quantum dot patterned 52\,nm away. The response of this detector is monotonic across the entire working voltage range of the device, which will make it particularly useful for studying systems of multiple quantum dots. The charge sensitivity is maximized when the junction is most conductive, suggesting that more sensitive detection can be achieved by shortening the length of the junction to increase its conductance.
\end{abstract}

\author{M. G. House}
\email{matthew.house@unsw.edu.au}
\affiliation{Centre for Quantum Computation and Communication Technology, University of New South Wales, Sydney, NSW 2052, Australia}
\author{E. Peretz}
\affiliation{Centre for Quantum Computation and Communication Technology, University of New South Wales, Sydney, NSW 2052, Australia}
\author{J. G. Keizer}
\affiliation{Centre for Quantum Computation and Communication Technology, University of New South Wales, Sydney, NSW 2052, Australia}
\author{S. J. Hile}
\affiliation{Centre for Quantum Computation and Communication Technology, University of New South Wales, Sydney, NSW 2052, Australia}
\author{M. Y. Simmons}
\email{michelle.simmons@unsw.edu.au}
\affiliation{Centre for Quantum Computation and Communication Technology, University of New South Wales, Sydney, NSW 2052, Australia}

\maketitle

The spin states of electrons in semiconductor quantum dots and donor sites have been an area of expanding research interest for the past decade due to their long quantum coherence times and applications to quantum information processing \cite{hanson07, zwanenburg13, loss98, kane98}. Quantum dots in semiconductor nanostructures were initially studied by measuring the transport of electrons through the quantum dots \cite{kouwenhoven97}. These studies were augmented by the development of nanoscale charge detection techniques, which allow the charge on a quantum dot to be measured by field effect and are commonly employed to study the electrostatics, excited state spectra, dynamics, and charge coherence of quantum dots \cite{field93, ihn09, gustavsson09}. Charge detectors can be used to measure electron spin states in single quantum dots by energy-dependent tunneling \cite{elzerman04} or by the Pauli blockade effect in a double quantum dot system \cite{petta05}.

Two types of charge sensor in wide use are the quantum point contact (QPC) \cite{field93} and the single-electron transistor (SET) \cite{barthel10}. These are field-effect devices in which the motion of a nearby charge changes the conductance of the channel significantly. SETs can be more sensitive than QPCs, in the sense of having a larger conductance change in response to one electron charge, but only at specific tunings where their Coulomb peaks occur \cite{barthel10}. QPCs have the advantage of operating over a wider range of gate voltages without requiring specific tuning. In silicon, conduction electrons have a short mean free path, so often the conductance of a QPC-like channel oscillates with respect to gate voltage due to coherent scattering effects, rather than exhibiting quantized conductance steps as in GaAs heterostructures \cite{house11, borselli11, nguyen13}. This behavior complicates charge detection because the response of the detector channel is non-monotonic and has ``blind spots'' where the sensitivity to small changes in the local potential is nil.

\begin{figure}
\begin{center}\includegraphics[width=6.5cm,keepaspectratio]{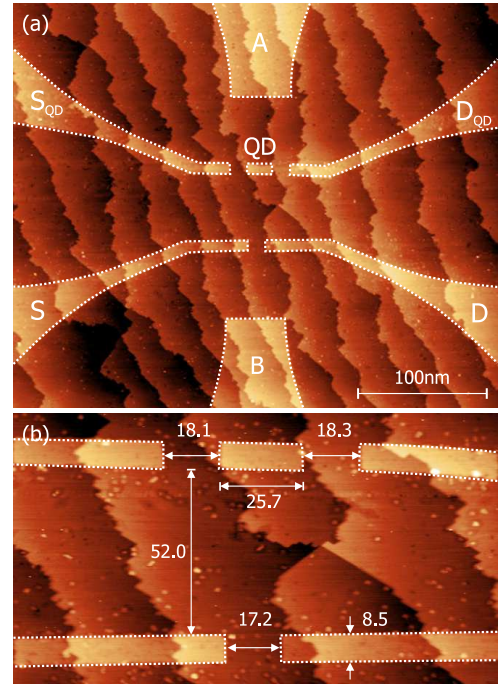}\end{center}
\caption{Tunnel junction charge sensor. (a) STM image of a tunnel junction charge sensor and a quantum dot whose charge is to be detected. Lighter colored areas show where the hydrogen mask has been removed and phosphorus dopants incorporate into the silicon. Leads S and D are separated by the tunnel junction. Leads S$_{\text{QD}}$ and D$_{\text{QD}}$ are tunnel-coupled to quantum dot QD. Gates A and B are designed to influence the potential of the quantum dot and the height of the tunnel barrier, respectively. (b) Detail of (a), with dimensions given in nm.}
\label{fig:stm_image}
\end{figure}

In highly doped, planar silicon devices fabricated by scanning tunneling microscope (STM) lithography, in-plane SETs have been used to detect the charge and spin states of quantum dots \cite{mahapatra11, buch13}. Although these detectors are very sensitive, the requirement to tune the SET to one of its sensitive points increases the complexity of the experiment in terms of the density of gates that must be patterned into the device and the complexity of voltage operations, \textit{e.g.} to perform a spin readout. The difficulty increases in devices with multiple quantum dots or spins to be read out, which makes it worthwhile to investigate an alternative to the SET. QPCs are difficult to implement in this system because conduction remains ohmic even when the width of the channel is reduced to only a few atoms \cite{weber12}. Instead, a field-effect device can be made by forming a short gap in a highly doped wire, which acts as a tunnel barrier for conduction electrons \cite{pokthesis, campbellthesis}. The electrostatic potential near the gap partially determines the height of the potential barrier, which in turn has an exponentially strong influence on the transmission of electrons. In this work we describe the design and fabrication of a sensitive field-effect transistor based on such a tunnel junction. The conductance of the junction responds to the field applied by a nearby gate and to single electron charging events on a quantum dot patterned nearby. The magnitude of the charge detection response is not only comparable to that of Si QPCs but also monotonic over a wide range of gate voltages. The sensitivity of the junction improves uniformly as its conductance increases, which suggests that the present results can be improved upon by shortening the junction to make it more conductive.

\begin{figure}
\begin{center}\includegraphics[width=\linewidth,keepaspectratio]{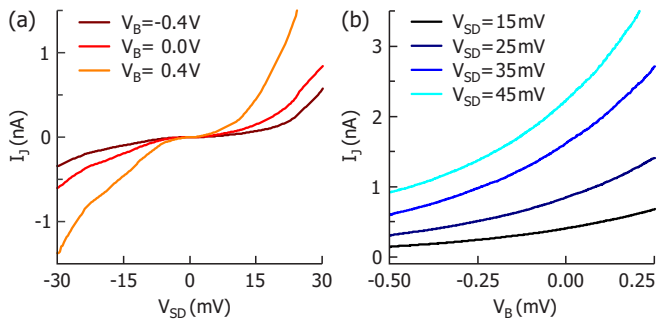}\end{center}
\caption{Characterization of the tunnel junction. (a) Current $I_{J}$ through the junction versus bias voltage $V_{SD}$ with $V_B=-0.4$ V (brown), $V_B=0.0$ V (red) and $V_B=0.4$ V (orange). (b) Dependence of the junction current on gate voltage $V_B$ for various bias values $V_{SD}$.}
\label{fig:quad_figure}
\end{figure}

The device is fabricated on a \textit{p}-type Si substrate (\mbox{1--10 $\Omega \cdot$cm}) in which the ($2 \times 1$) surface reconstruction is prepared in ultra-high vacuum by heating the sample to 1100\,$^\circ$C followed by a controlled cool-down at rate of ~5$\,^\circ$C\,$\cdot$\,s$^{-1}$ to 330\,$^\circ$C.  The surface is then terminated with monoatomic hydrogen, which is selectively removed with the STM tip to create a mask for subsequent adsorption of gaseous PH$_3$ precursor molecules onto the surface \cite{schofield03, fuechsle12}. Next the phosphorus atoms are incorporated into the silicon crystal by a 60\,s anneal to 330 $^\circ$C, and finally encapsulated with 31\,nm of epitaxial silicon. This results in a substrate which is insulating at low temperatures, containing highly doped metallic features patterned with atomic precision. The hydrogen mask of the device is presented in Fig.~1. Leads S and D are separated by a gap in the doping profile 17.2\,nm long and 8.5\,nm wide, which acts as a tunnel junction. A quantum dot (labelled QD in Fig. \ref{fig:stm_image}) is patterned 52\,nm away, contacted by tunnel barriers to leads through which the conductance of the quantum dot are measured. Two gates A and B allow tuning of the tunnel junction and of the quantum dot potential.

Fig.~\ref{fig:quad_figure}(a) shows that the current $I_J$ through the junction depends nearly exponentially on the applied bias $V_{SD}$, with some variation due to nonuniform density of states in the leads. These curves demonstrate how the conductance can be changed by the potential on gate B. The zero-bias resistance of the junction is 1.0\,G$\Omega$ with $V_B=0$ V, and can be tuned from 3\,G$\Omega$ at $V_B=-0.8$\,V to $160$\,M$\Omega$ at $V_B=+0.8$\,V. The influence of gate B on the junction conductance at finite bias is shown in Fig.~2(b). The response is exponential, as expected for tunneling through a barrier of variable height, with no evidence of conductance oscillations due to disorder in the barrier. We note that gate A influences the junction conductance in a way similar to gate B but the influence is weaker due to gate A being further away from the junction and partially screened by the quantum dot channel.

\begin{figure}
\begin{center}\includegraphics[width=\linewidth,keepaspectratio]{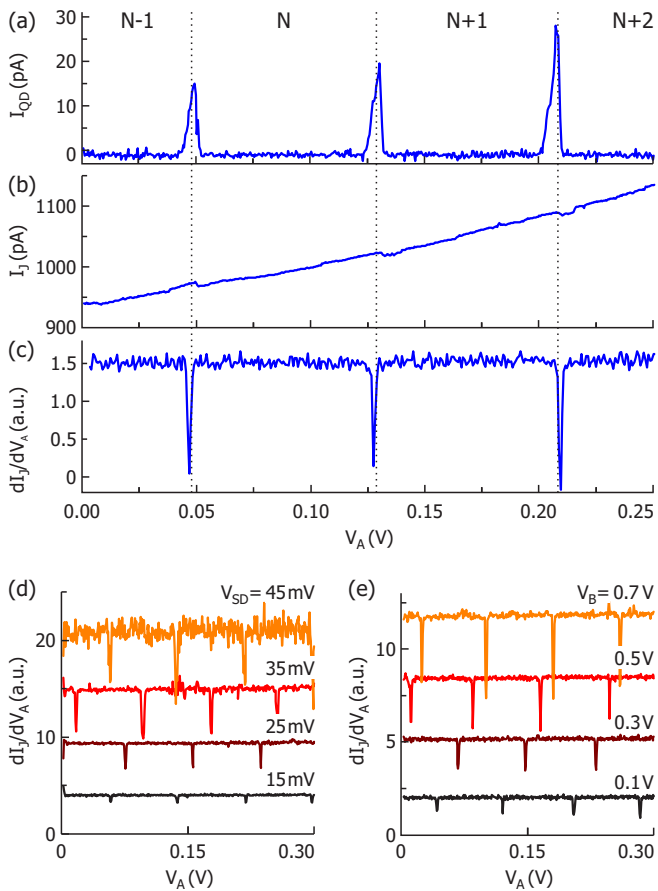}\end{center}
\caption{Response of the junction conductance to charging events. (a) Current $I_{QD}$ through the quantum dot as a function of gate voltage $V_A$, with a 1\,mV bias applied. Three Coulomb blockade peaks indicate three electron transitions in this range of gate voltage. (b) Current $I_{J}$ through the tunnel junction. Dips in the current correspond with the Coulomb peaks of the quantum dot. (c) Transconductance $dI_{J}/dV_A$ of the junction with respect to the voltage on gate A. Changes in the junction conductance due to electron transitions of the quantum dot apear as sharp peaks. (d) Change of the charge detection peaks due to changing bias voltage $V_{SD}$, with $V_B=0$\,V. Increasing the bias increases the charge detection amplitude, although it also increases noise in the measurement at high bias. (e) Change of the charge detection peaks due to gate voltage $V_B$, with $V_{SD}=20$ mV. The peak amplitude increases with increasing $V_B$, as the junction conductance increases.}
\label{fig:tri_figure}
\end{figure}

The conductance of the gap also responds to changes in the charge configuration of the quantum dot. \mbox{Fig.~\ref{fig:tri_figure}(a)} shows a measurement of the current through the quantum dot as a function of gate voltage $V_A$. There are a series of Coulomb peaks, each of which indicates a change in the number of electrons on the quantum dot, as indicated by $N-1$, $N$, ... on the plot. The current through the junction, Fig.~\ref{fig:tri_figure}(b), shows dips that are coincident with the Coulomb peaks, demonstrating that the conductance of the junction is influenced by the charge transitions of the quantum dot. To emphasise these features and reduce 1/$f$ noise, which is the dominant noise source in \mbox{Fig. \ref{fig:tri_figure}(b)}, we use a transconductance measurement by applying an a.c. signal (0.8\,mV RMS amplitude, 19.43\,Hz) to gate A and measuring the resulting modulation of $I_J$ with a lock-in amplifier. The result of this measurement is shown in Fig.~\ref{fig:tri_figure}(c). The clear peaks in the transconductance correspond with the Coulomb peaks of the quantum dot, with an improved signal-to-noise ratio compared to the d.c. measurement.

\begin{figure}
\begin{center}\includegraphics[width=\linewidth,keepaspectratio]{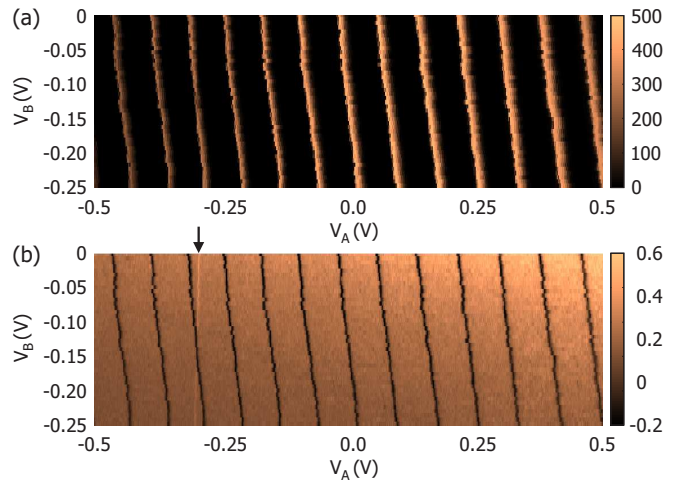}\end{center}
\caption{Charge stability diagram. Comparison of (a) the current through the quantum dot $I_{QD}$ (in pA) and (b) the a.c. current response of the tunnel junction channel (in arbitrary units) as a function of the two gate voltages $V_A$ and $V_B$. The bias on quantum dot is 1\,mV and on the junction 10\,mV. Peaks in the AC current of the junction correspond with Coulomb peaks of the quantum dot, demonstrating that they detect individual electron transitions of the ground state of the quantum dot. The arrow indicates an additional charge trap feature, not associated with the quantum dot, which was also detected in this device.}
\label{fig:stability_diagram}
\end{figure}

Fig. \ref{fig:stability_diagram} shows a comparison of the quantum dot conductance, Fig.~\ref{fig:stability_diagram}(a), with that of the junction transconductance, Fig.~\ref{fig:stability_diagram}(b), as a function of the two gate voltages $V_A$ and $V_B$. Both reveal straight lines, indicative of a single quantum dot (with charging energy 8\,meV, as determined by a separate Coulomb diamond measurement). We see that the charge detection signal is monotonic and nearly equal in strength across the entire range of gate voltages. This makes the tunnel junction charge sensor especially useful for detecting charge transitions in multiple-dot or multiple-donor devices with complex stability diagrams. The increased range of sensitivity helps to detect an additional feature, indicated by an arrow in \mbox{Fig. \ref{fig:stability_diagram}(b)}, due to a charge transition of a defect state tunnel-coupled to gate A.

The magnitude of the charge detection signal, evaluated either as the height of a step in the conductance, [Fig. \ref{fig:tri_figure}(b)], or as the magnitude of a transconductance peak [Fig. \ref{fig:tri_figure}(c)], increases uniformly with the conductance of the junction. The conductance can be influenced by both the junction bias $V_{SD}$ and by $V_B$. \mbox{Fig. \ref{fig:tri_figure}(d)} shows the transconductance peaks for four different biases. The peaks increase in magnitude along with bias; at the highest biases an increase in the noise level limits the increase of the signal-to-noise ratio. Similarly, Fig.~\ref{fig:tri_figure}(e) shows an increase in the detection signal as the junction is tuned to be more conductive by making $V_B$ more positive. $V_B$ can be increased only until a significant leakage current begins to flow from the gate to the channel (the working range is $|V_A|,|V_B|<\pm 0.8$\,V in this device). Both of these observations indicate that the charge sensitivity of the junction could be improved by fabricating a shorter junction to increase its conductance. The charge detection sensitivity demonstrated in this experiment is $\sim10^{-2}$ e/$\sqrt{\text{Hz}}$, limited by noise in the room-temperature current amplifier. The same signal amplitude would correspond to a sensitivity of $\sim10^{-3}$ e/$\sqrt{\text{Hz}}$ if noise in the measurements were at the theoretical shot noise limit $\sqrt{2 e \langle I\rangle \Delta f}$. The fractional change in conductance due to a single charge is comparable to other silicon charge detectors  \cite{house11, borselli11, nguyen13}.

In summary, we detected single electron charges using a planar, nanometer-scale tunnel junction fabricated in silicon by STM lithography. The conductance of this junction responds to the electrostatic field of a gate and to electron transitions of a quantum dot. The sensitivity we demonstrated is similar to that of QPC charge detectors in Si, but the response is monotonic over a wide range of gate voltages which means it requires no special operation to maintain it at a point of maximum sensitivity. A uniform response will be useful in particular for investigating multiple-dot or multiple-donor devices with complex stability diagrams. The fractional change in conductance due to a single charge is nearly constant regardless of the device tuning, so the charge detection sensitivity improves as the junction is more conductive. The conductance of such a junction can be engineered over many orders of magnitude depending on the length and width of the gap \cite{pokthesis}, so the sensitivity can be improved in future devices by fabricating a junction with smaller tunnelling distance. 

This research was supported by the Australian Research Council Centre of Excellence for Quantum Computation and Communication Technology (project number CE110001027), the U.S. National Security Agency and the U.S. Army Research Office under contract number W911NF-13-1-0024. M.Y.S. acknowledges an Australian Research Council Laureate Fellowship.

\bibliographystyle{apsrev4-1} 

\end{document}